\documentclass[twocolumn,aps,superscriptaddress,prb]{revtex4-1}
\usepackage{graphicx}
\usepackage{amssymb}
\usepackage{amsmath}
\usepackage{bm}
\usepackage{color}

\usepackage{braket}
\usepackage{siunitx}
\usepackage[top=30truemm,bottom=30truemm,left=25truemm,right=25truemm]{geometry}

\def\Vec{\mathbf}

\def\lsim{\, \lower -0.3ex \hbox{$<$} \kern -0.75em \lower 0.7ex \hbox{$\sim$} \,}
\def\gsim{\, \lower -0.3ex \hbox{$>$} \kern -0.75em \lower 0.7ex \hbox{$\sim$} \,}

\begin{document}

\title{
Moir\'{e} edge states in twisted bilayer graphene and their topological relation to quantum pumping
}
\author{Manato Fujimoto}
\affiliation{Department of Physics, Osaka University,  Osaka 560-0043, Japan}
\author{Mikito Koshino}
\affiliation{Department of Physics, Osaka University,  Osaka 560-0043, Japan}
\date{\today}

\begin{abstract} 

We study the edge states of twisted bilayer graphene and their topological origin.
We show that the twisted bilayer graphene has special edge states associated with the moir\'{e} pattern,
and the emergence of these moir\'{e} edge states is linked with the sliding Chern number, 
which describes topological charge pumping caused by relative interlayer sliding. 
When one layer of the twisted bilayer is relatively slid with respect to the other layer, 
the edge states are transferred from a single band to another
across the band gap, and the number of the edge states pumped in a sliding cycle is shown to be 
equal to the sliding Chern number of the band gap. 
The relationship can be viewed as a manifestation of the bulk-edge correspondence inherent in moir\'{e} bilayer systems.

\end{abstract}

\maketitle

%%%%

\section{Introduction}

In condensed matter systems, the topology of electronic  band is intimately related to the emergence of edge states, i.e.,
electronic states localized at the boundary of the system.\cite{PhysRevLett.49.405,kane2005z,qi2008topological}
In general, the existence of edge states in a specific band gap is related to the non-zero topological invariant in the bulk system. \cite{PhysRevLett.71.3697,PhysRevB.48.11851}
In quantum Hall systems, for example, the number of edge modes coincides with the summation of the Chern numbers over all the occupied bands below the gap.
\cite{PhysRevLett.49.405,KOHMOTO1985343, PhysRevLett.71.3697,PhysRevB.48.11851} 
Similar relationships between bulk topological property and emergent edge modes are found in a wide variety of physical systems, including 
topological insulators, topological superconductors,\cite{hasan2010colloquium,qi2011topological} mechanical systems\cite{kane2014topological,kariyado2015manipulation,sstrunkE4767} and photonic systems.\cite{RevModPhys.91.015006,lu2014topological}

%The topology of energy bands plays an important role to explain the existence of edge state which is localized at the boundary of condensed matter systems.\cite{PhysRevLett.49.405,kane2005z,qi2008topological,hasan2010colloquium,qi2011topological}
%To give examples, a topological invariant, Chern number, characterizes the number of edge state in integer quantum Hall effect\cite{PhysRevLett.49.405,KOHMOTO1985343, PhysRevLett.71.3697,PhysRevB.48.11851} and psuedospin winding number also characterizes that of monolayer graphene truncated by zigzag boundary.\cite{PhysRevLett.89.077002}
%As described in examples, in general, the existence of edge states in a specific band gap is related to the non trivial topological invariant of all the bulk bands below the band gap.\cite{PhysRevLett.71.3697,PhysRevB.48.11851,hatsugai2016bulk}
%The relationship is called “bulk-edge correspondence”.

In this paper, we study the edge states of twisted bilayer graphenes (TBG) 
and their topological origin.
TBG is a two-dimensional material where two graphene layers are overlapped with an arbitrary twist angle.
In a low-angle TBG, the long-range moir\'{e} pattern strongly modifies the graphene's Dirac cone\cite{PhysRevLett.99.256802,PhysRevB.81.161405,trambly2010localization,PhysRevB.81.165105,PhysRevB.82.121407,bistritzer2011moire,PhysRevB.83.045425,PhysRevB.84.075425,dos2012continuum,PhysRevB.85.195458,PhysRevB.86.125413,PhysRevB.87.205404}, resulting in a flat band at zero energy.\cite{bistritzer2011moire,PhysRevB.86.125413,PhysRevLett.117.116804,cao2018unconventional}
Some previous works studied the edge properties of TBG,\cite{PhysRevB.87.075433,PhysRevB.89.205405,PhysRevB.91.035441,pelc2015electronic,fleischmann2018moire,liu2019pseudo}
and it was shown that TBG has two kinds of the edge states\cite{fleischmann2018moire}:
One is zero-energy edge modes on the zigzag termination \cite{PhysRevB.87.075433,PhysRevB.89.205405,PhysRevB.91.035441,pelc2015electronic}, which are inherited from monolayer graphene. \cite{PhysRevB.59.8271}
The other one, which we refer to moir\'{e} edge state, is
qualitatively different state strongly dependent on the moir\'{e} pattern,
and occurs away from zero energy. \cite{fleischmann2018moire, liu2019pseudo}
Around the magic angle, in particular, the moir\'{e} edge states
come to the energy gaps between flat band and excited band.\cite{liu2019pseudo}
%However, the bulk-edge correspondence of the non-zero edge states is not unveiled.

% In this paper, we study the edge state of the twisted bilayer graphenes (TBG).
%TBG is a two-dimensional material where two graphene layers are relatively rotated by an arbitrary angle.
%The system in low-angle has the flat-band\cite{bistritzer2011moire,PhysRevB.86.125413} which enhances strongly correlated electron-electron interactions.\cite{PhysRevLett.117.116804,cao2018unconventional,cao2018correlated,nam2017lattice,koshino2018maximally}
%Some works studied the edge states of TBG.\cite{PhysRevB.87.075433, PhysRevB.89.205405,PhysRevB.91.035441,fleischmann2018moire,liu2019pseudo}
%TBG nanoribbon possesses localized zero energy edge states found on the zigzag boundary,\cite{PhysRevB.59.8271} and also, non zero energy edge states which is strongly modulated by the moir\'{e} pattern.\cite{fleischmann2018moire}
%Especially, around the magic angle, the non-zero edge states in the energy gaps between flat band and excited band.\cite{liu2019pseudo}
%However, the bulk-edge correspondence of the non-zero edge states is not unveiled.

One may ask if the moir\'{e} edge states are related to some sort of bulk topology.
The energy bands of TBG has zero Chern number, and hence the system does not have any edge states
associated with the Hall effect or the valley Hall effect.
Recent works \cite{PhysRevB.101.041112,zhang2020topological, su2020topological} proposed a different topological invariant for TBG, called sliding Chern number, 
which represents the number of adiabatic charge pumping\cite{PhysRevB.27.6083} caused by a mechanical interlayer sliding. 
More specifically, when one layer of TBG is relatively slid with respect to the other layer by a single atomic period,
then electrons on the TBG are pumped by an integer multiple of the moir\'{e} period, where the integer is given by the sliding Chern number.

Here we investigate the edge states of TBG under the effect of the interlayer sliding,
and find that the emergence of the moir\'{e} edge states is linked with the nonzero sliding Chern numbers.
We demonstrate that the edge states are transferred in the energy axis from the flat band to the excited band during the interlayer sliding process,
and the number of edge states pumped in a sliding cycle is equal to the sliding Chern number of the band gap.
The relationship can be viewed as a bulk-edge correspondence inherent in moir\'{e} bilayer systems.
%We also show that the edge states split off from the bulk energy band right when the boundary crosses the AA-spot.

% The flat-band is predicted to exhibit a fragile topology\cite{po2019faithful,ahn2019failure,song2019all,carr2019derivation, song2020twisted} in which the gap do not have non-zero valley Chern numbers and symmetry protected gapless surface. 
%On the other hand, in our previous work\cite{PhysRevB.101.041112} we proposed a new topological number, called sliding Chern number, which represents the number of adiabatic charge pumping caused by interlayer sliding.
%To understand the edge state of TBG, study of the bulk-edge correspondence of non-zero sliding Chern number is needed. 
%In this paper, we study the relationship between the interlayer sliding and the electric structure of TBG nanoribbon.
%We show that the number of edge states in one cycle of sliding is equal to the sliding Chern number by calculating band structure of TBG nanoribbon.

The paper is organized as follows. 
In Sec.\ II, we introduce the atomic structure of TBG and the tight binding model. 
In Sec.\ III, we calculate the energy spectrum of TBG nanoribbon as a function of sliding distance,
and demonstrate the correspondence between the edge states and the  sliding Chern number.
A brief conclusion is given in Sec.\ IV.

%%%
%%%%%%%%%%%%%%%%%%%%%%%%%%%
\section{Model}
\label{sec_model}
\subsection{Atomic structure}
\label{sec_atom_structure}

\begin{figure}
  \begin{center}
    \leavevmode\includegraphics[width=0.8 \hsize]{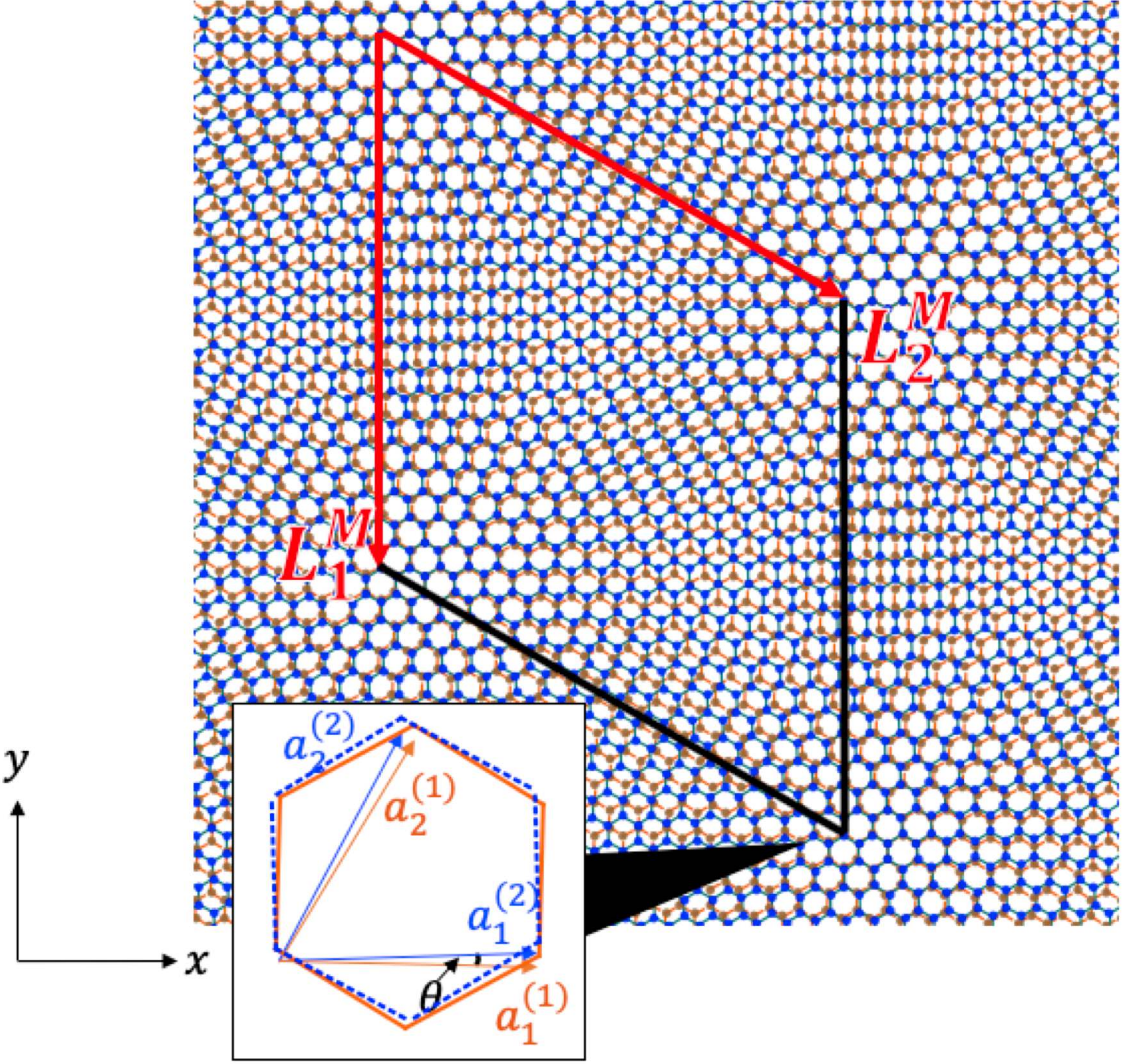}
 \caption{Atomic structures and the moir\'{e} unit cell of TBG with $\theta=3.15^\circ$. The inset shows the primitive lattice vectors of layers 1 and 2.}
    \label{fig_moire_atomic_structure}
  \end{center}
  \end{figure}
 
TBG can be generated from AA-stacked bilayer graphene (i.e., perfectly overlapping
honeycomb lattices) by rotating layer 1 and 2 by $-\theta/2$
and $\theta/2$, respectively.
Figure \ref{fig_moire_atomic_structure} illustrates the atomic structure of TBG of $\theta=3.15^{\circ}$.
We define $\mathbf{a}_{1}=a(1,0), \mathbf{a}_{2}=a(1 / 2, \sqrt{3} / 2)$ as the lattice vectors of monolayer graphene
before the rotation, where $a=0.246 \mathrm{nm}$ is graphene's lattice constant. 
%The corresponding reciprocal lattice vectors are $\mathbf{a}_{1}^{*}=(2 \pi / a)(1,-1 / \sqrt{3})$ and $\mathbf{a}_{1}^{*}=(2 \pi / a)(0,2/\sqrt{3})$.
The lattice vectors of layer $l$ after the rotation are given by $\mathbf{a}_{i}^{(l)}=R(\mp \theta / 2) \mathbf{a}_{i}$ with $\mp$ for  $l=1,2$ respectively, where $R(\theta)$ represents the rotation by $\theta$ on $xy$-plane. 
%Likewise, the reciprocal lattice vectors become $\mathbf{a}_{i}^{*(l)}=R(\mp \theta / 2) \mathbf{a}_{i}^{*}$.

When the rotation angle is small, the mismatch between the lattice structures of the two layers gives rise to a long-range moir\'{e} pattern, which is ruled by the primitive lattice vectors,
\begin{align}
\mathbf{L}_{i}^{M}&=[R(\theta / 2)-R(-\theta / 2)]^{-1} \mathbf{a}_{i} \quad (i=1,2) \nonumber\\
&=\frac{1}{2 \sin (\theta / 2)}R(-\pi / 2) \mathbf{a}_{i},
%\nonumber\\
%&=\frac{1}{2 \sin (\theta / 2)}R(-\pi / 2 \mp \theta/2) \mathbf{a}_{i}^{(l)}
\end{align}
with $\mp$ for  $l=1,2$.
The lattice constant $L_{\mathrm{M}}=\left|\mathbf{L}_{1}^{\mathrm{M}}\right|=\left|\mathbf{L}_{2}^{\mathrm{M}}\right|$ is given by
\begin{equation}
L_{M}=\frac{a}{2 \sin (\theta / 2)}.
\end{equation}

%In Fig.\ref{fig_moire_atomic_structure}, the corners of the unit cell correspond to the AA spots, where
%the honeycomb lattices of layer 1 and 2 are  overlap. 

 \begin{figure}
  \begin{center}
    \leavevmode\includegraphics[clip,width=1. \hsize]{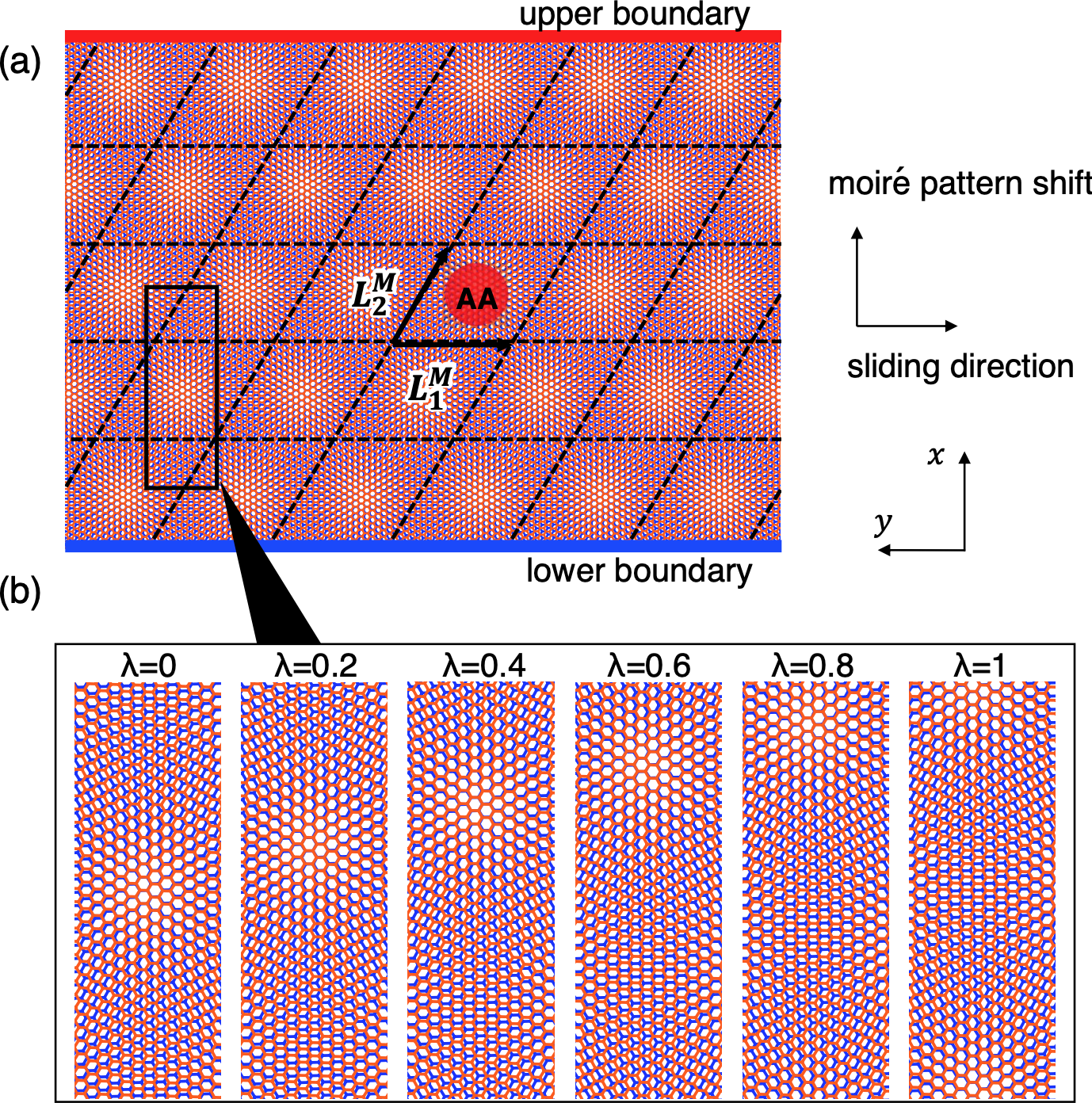}
    \caption{(a) TBG nanoribbon with $\theta=3.15^\circ$ truncated by upper (red) and lower (blue) boundaries.
   Note that the figure is rotated by 90$^\circ$ so that $x$ axis is vertical and $y$ axis is horizontal.
The structure is periodic in the direction of $\mathbf{L}_{1}^M$, and five unit cells thick in the perpendicular direction.
(b) Detailed atomic structure in the sliding parameter $\lambda=0,0.2,...,1$.
}
    \label{fig_TBG_nanoribbon_structure}
  \end{center}
  \end{figure}  
 
 % interlayer sliding and moire movement
 
If we slide the layer $l$ by $\mathbf{a}_{i}^{(l)}$ with the other layer fixed, 
the moir\'{e} pattern shifts exactly by $\pm \mathbf{L}_{i}^{M}$ for $l=1,2$, respectively. %\cite{appendix}
Therefore, when we slide the layer $l$ by an arbitrary displacement vector, 
\begin{equation}
\label{eq_slide_ls}
\Delta \mathbf{x}^{(l)}=\nu_1 \mathbf{a}_{1}^{(l)} + \nu_2 \mathbf{a}_{2}^{(l)}, 	
\end{equation}
then the moir\'{e} pattern moves by
\begin{align}
\label{eq_slide_ls2}
\Delta \mathbf{X}&=\pm(\nu_1 \mathbf{L}_{1}^{M} + \nu_2 \mathbf{L}_{2}^{M}) \nonumber\\
&=\frac{\pm1}{2 \sin (\theta / 2)}R(-\pi / 2 \mp \theta/2) \Delta\mathbf{x}^{(l)},
\end{align}
where each double sign corresponds to $l=1,2$, respectively.
When $\theta \ll 1$,  the moir\'{e} pattern shift $\Delta \mathbf{X}$ is nearly perpendicular to the sliding vector $\Delta\mathbf{x}^{(l)}$,
and its amplitude is magnified by the factor $[2 \sin (\theta / 2)]^{-1} \sim 1/\theta$.

 % ribbon geomtery
In the following, we consider a TBG nanoribbon as shown in Fig.\ \ref{fig_TBG_nanoribbon_structure}(a) to investigate the edge states. 
Note that the figure is rotated by 90$^\circ$ so that $x$ axis is vertical and $y$ axis is horizontal.
Here we assume that the ribbon is parallel to $y$ and five unit cells thick in the perpendicular direction  (along $x$),
truncated by red and blue lines.
The boundary is nearly parallel to the armchair direction of graphene, 
so that the zigzag edge states of monolayer graphene are almost absent.

Now we slide layer 2 with respect to layer 1 in along the length of the ribbon ($y$),
to move the moir\'{e} pattern along the width ($x$).
We specify the sliding vector by $(\nu_1,\nu_2) = \lambda (1/2, -1)\, (0\leq \lambda \leq 1)$,
which gives $\Delta \mathbf{x}^{(2)}$  almost along $-y$ direction.
When the sliding parameter $\lambda$ is changed from 0 to 1, 
the moir\'{e} pattern moves by $\Delta \mathbf{X} = -(1/2) \mathbf{L}_{1}^{M}+ \mathbf{L}_{2}^{M}$ exactly in the $x$-direction.
After the process, all the AA spots move just by one row as illustrated in the lower panel of Fig.\ \ref{fig_TBG_nanoribbon_structure}(b).
Because of the triangular-lattice arrangement, the AA spots of do not come back to the original positions, 
but as we will see, this process virtually gives a single cycle of the edge states pumping.

%%%%%%%%%%%
\subsection{Tight binding model}
\label{sec_tbmodel}

We calculate the eigenenergies and the eigenfunctions of the TBG ribbon using the tight-binding model for carbon $p_z$ orbitals.
The Hamiltonian is written as  \cite{trambly2010localization,nakanishi2001conductance,PhysRevB.69.075402,PhysRev.94.1498}
\begin{equation}
 \label{eq_Hamiltonian_1D}
H=-\sum_{\langle i, j\rangle} t\left(\mathbf{R}_{i}-\mathbf{R}_{j}\right)\left|\mathbf{R}_{i}\right\rangle\left\langle \mathbf{R}_{j}\right|+\mathrm{H.c.}
\end{equation}
where $\mathbf{R}_i$ and $\left|\mathbf{R}_{i}\right\rangle$ represent the lattice point and the atomic state at site $i$, respectively, and $ t\left(\mathbf{R}_{i}-\mathbf{R}_{j}\right)$ is the transfer integral between site $i$ and site $j$.
We adopt the Slater-Koster-type formula for the transfer integral,
\begin{equation}
\begin{aligned}-t(\mathbf{d}) &=V_{p p \pi}\left[1-\left(\frac{\mathbf{d} \cdot \mathbf{e}_{z}}{d}\right)^{2}\right]+V_{p p \sigma}\left(\frac{\mathbf{d} \cdot \mathbf{e}_{z}}{d}\right)^{2} \\ V_{p p \pi} &=V_{p p \pi}^{0} \exp \left(-\frac{d-a_{0}}{\delta_{0}}\right) \\ V_{p p \sigma} &=V_{p p \sigma}^{0} \exp \left(-\frac{d-d_{0}}{\delta_{0}}\right) \end{aligned},
\label{eq_hopping}
\end{equation}
where $\mathbf{d}=\mathbf{R}_{i}-\mathbf{R}_{j}$ is the distance between two atoms, and $\mathbf{e}_{z}$ is the  unit vector on the $z$ axis. 
$V_{p p \pi}^0$ is the transfer integral between the nearest-neighbor atoms of monolayer graphene which are located at distance $a_{0}=a / \sqrt{3} \approx 0.142$ nm.
$V_{p p \sigma}^0$ is the transfer integral between two nearest vertically aligned atoms, $d_{0} \approx 0.335 \mathrm{nm}$ is the interlayer spacing. 
$\delta_0$ is decay length of the transfer integral and is chosen as $0.184a$.
The transfer integral for $d > \sqrt{3} a$ is exponentially small and can be safely neglected.

Our target is the TBG of the magic angle ($\theta = 1.05^\circ$) which has nearly flat bands. \cite{bistritzer2011moire,PhysRevB.86.125413,PhysRevLett.117.116804,cao2018unconventional,cao2018correlated,nam2017lattice,koshino2018maximally}
It has a huge number of atoms (about 12,000) in the moir\'{e} unit cell, requiring a large computational cost to calculate the energy bands of the ribbon
in the tight-binding model.
To reduce the number of atoms, we take the atomic structure of $\theta=3.15^\circ$ which has 10 times fewer atoms per unit cell,
but at the same time, enlarge the interlayer hopping energy (i.e., transfer integral between atoms of layer 1 and layer 2)
by factor of $3.15^{\circ} / 1.05^{\circ} = 3$,
to mimic the band structure of $\theta=1.05^\circ$.
The approximation works for the follow reason. The low-energy band structure of TBG is determined by the ratio of two energy scales, $t_{\rm inter}/E_M$,
where $t_{\rm inter}$ is the interlayer hopping energy,
and $E_M =\hbar v (2\pi/ L_M)\sim (2\pi \hbar v /a)\theta$ is the moir\'{e} band folding energy with graphene's band velocity $v$.\cite{PhysRevLett.99.256802,PhysRevB.83.045425,bistritzer2011moire,PhysRevB.87.205404,dos2012continuum,koshino2015interlayer,koshino2015electronic,weckbecker2016low}
Therefore, the TBG of $\theta=3.15^\circ$ with $t_{\rm inter}$ multiplied by factor 3
has a nearly identical low-energy band structure as the original $\theta \sim 1.05^\circ$ model except for the overall energy scale.
%To reproduce the band structure of $\theta = 1.05^\circ$, we calculate the tight-binding model of $\theta=3.15^\circ$
%with $t_{\rm inter}$ multiplied by $3.15^{\circ} / 1.05^{\circ} = 3$, to have the same $t_{\rm inter}/E_M$ as the original $\theta \sim 1.05^\circ$ model.
Also, we include the in-plane lattice relaxation, which gives an energy gap between flat bands and the excited bands\cite{nam2017lattice}.
Here we set the lattice displacement vector ${\bf u}$ that is equal to that of $\theta=1.05^{\circ}$.

\begin{figure}
  \begin{center}
    \leavevmode\includegraphics[width=1. \hsize]{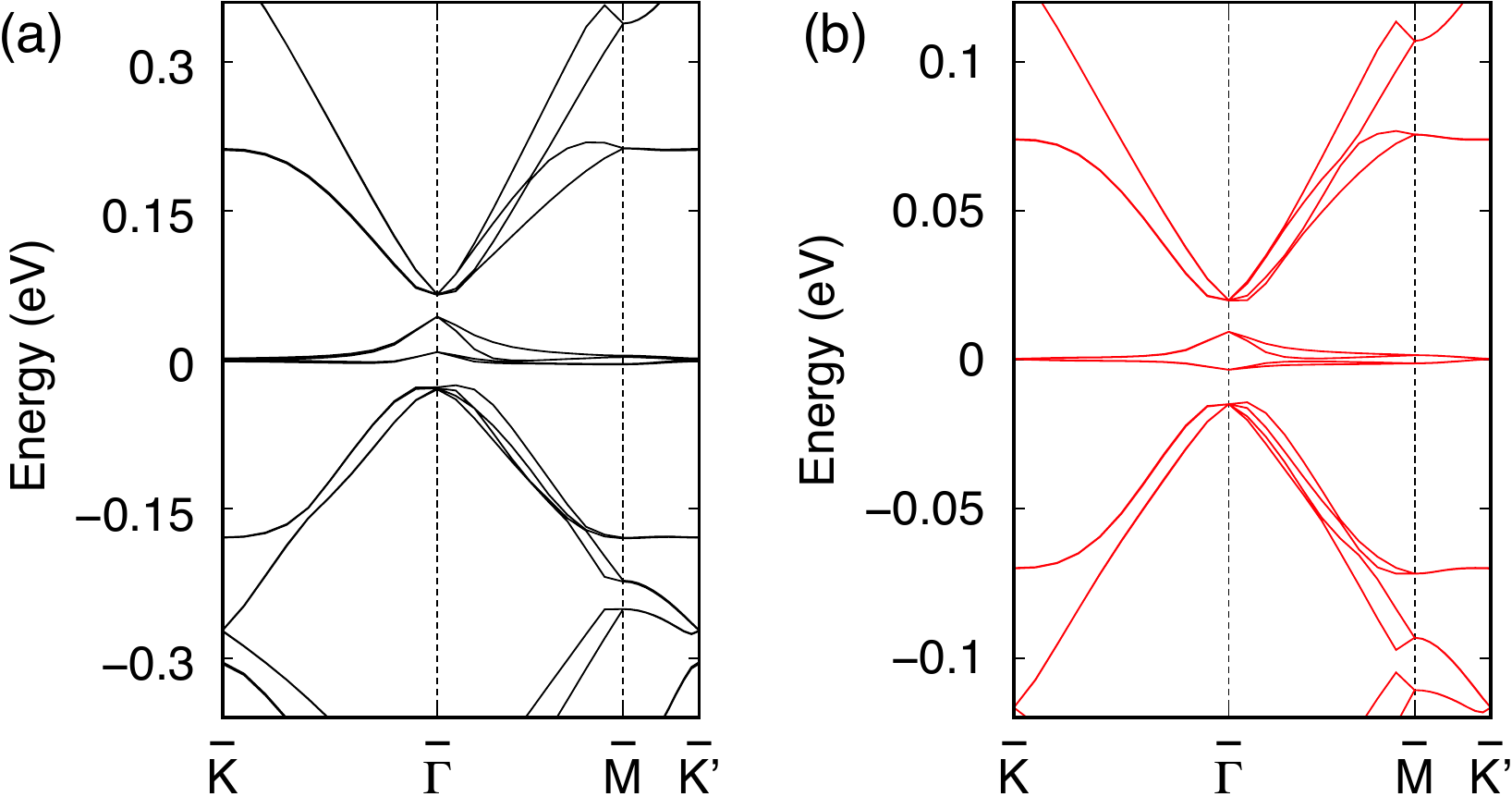}
    \caption{Band structures of (a) $\theta=3.15^\circ$ tight-binding model with enlarged interlayer hopping, 
and (b) of the original $\theta=1.05^\circ$ model (see the text).}
    \label{fig_band_model}
  \end{center}
  \end{figure}

Figure \ref{fig_band_model} plots the bulk band structure calculated for (a) the $\theta=3.15^\circ$ model with three-times enlarged interlayer hopping, 
and (b) the original $\theta=1.05^\circ$ model, showing a nice qualitative agreement except for the energy scale difference by the factor of 3.
In the following calculation, the energy scale is shrunk by 1/3 times to emulate $1.05^{\circ}$ TBG.
Note that the TBG of $\theta=3.15^\circ$ is a commensurate system
at which the atomic structure is exactly periodic in the moir\'{e} period $\mathbf{L}_{i}^M$, \cite{PhysRevLett.99.256802,PhysRevB.83.045425,bistritzer2011moire,PhysRevB.87.205404,dos2012continuum,koshino2015interlayer,koshino2015electronic,weckbecker2016low}
and hence the eigenstates of the tight-binding model can be obtained by diagonalizing a finite-sized Hamiltonian matrix.

%%%%%%%%%%%%%
\section{Result}
\label{sec_result}
\subsection{Electronic structure and edge states}
\label{sec_edge}

 \begin{figure*}[t]
 \begin{center}
  \includegraphics[width=1. \hsize]{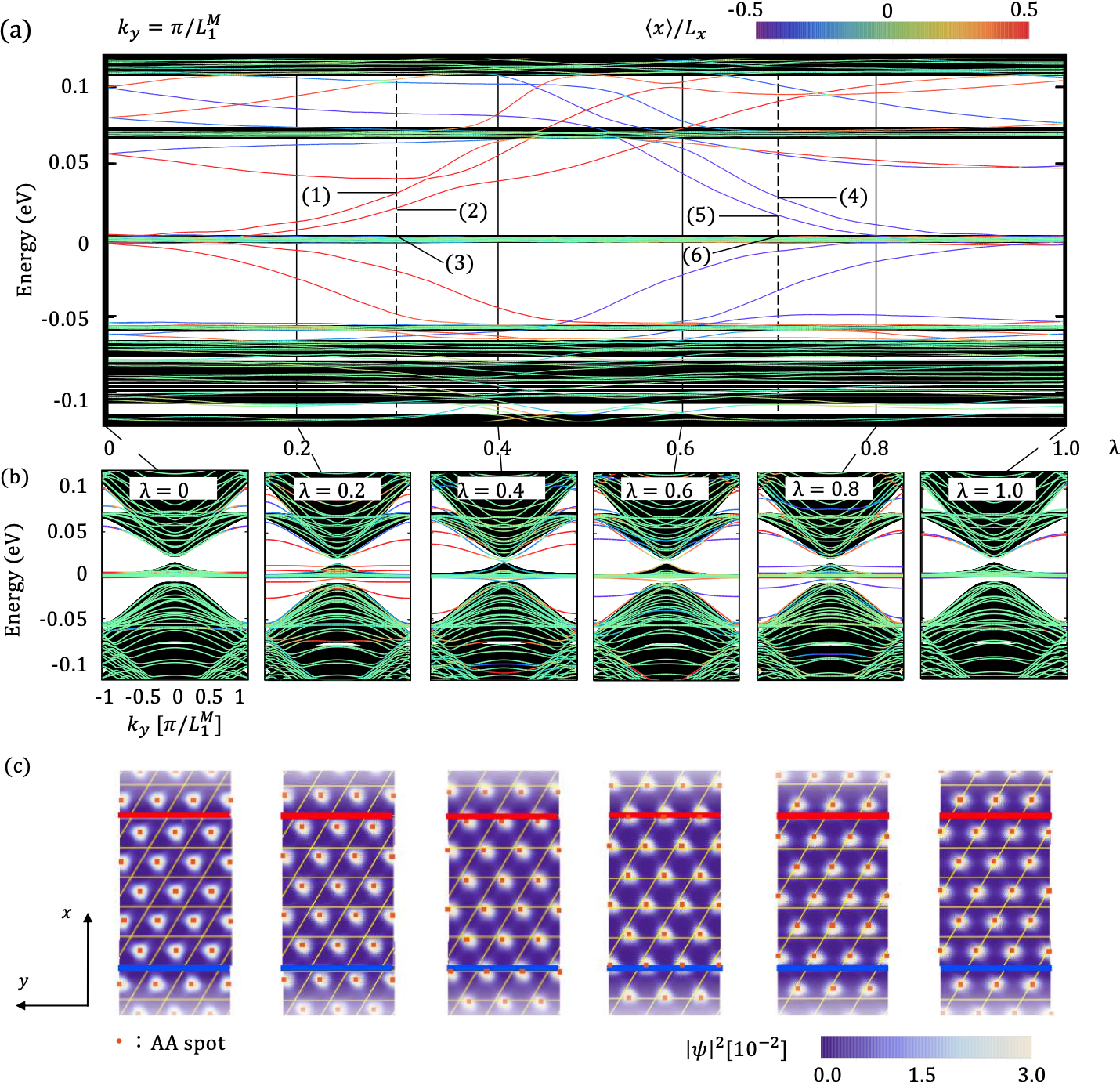}
  \caption{
  (a) Energy band structure of TBG nanoribbon at $k_y=\pi/L_1^M$ as a function of sliding parameter $\lambda$,
and (b) the full band structure against $k_y$ at $\lambda = 0, 0.2, \cdots, 1$.
Here the colored lines are the energy bands of TBG ribbons, where the color represents the expected value of $x$ coordinate,
and the black-colored areas in background represent the energy bands of the bulk TBG projected onto $k_y$ axis.
(c) Real-space map of the flat-band wave function at $\bar{K}$ of bulk TBG 
at $\lambda = 0, 0.2, \cdots, 1$ [corresponding to upper panels in (b)]. The red and blue lines represent the boundary lines for the ribbon.}
  \label{fig_ribbon_band_vs_slide}
 \end{center}
\end{figure*}

The calculated band diagram of the TBG ribbon is summarized in Fig.\ \ref{fig_ribbon_band_vs_slide}.
Here Fig.\ \ref{fig_ribbon_band_vs_slide}(a) shows the band energies at the fixed wave number $k_y = \pi / L_1^M$,
as a function of sliding parameter $\lambda$.
The panel (b) presents the full band structure against $k_y$, at different sliding distances $\lambda = 0, 0.2, \cdots, 1$.
In Fig.\ \ref{fig_ribbon_band_vs_slide} (a) and (b), 
the colored lines are the energy bands of TBG ribbons, where the color represents the expected value of $x$ coordinate;
red (blue) lines indicate edge states localized at the upper (lower) boundary, while green lines are bulk states spreading over the entire system.
The black-colored areas in background represent the energy bands of the bulk TBG projected onto $k_y$ axis.

In increasing $\lambda$ from 0 to 1, we see that two edge bands of the upper boundary (red lines) split off from the zero-energy flat band 
in each of positive and negative energy sides,
and they are eventually absorbed into the excited conduction/valence bands.
At the same time, two edge bands of the lower boundary (blue lines) transfer from the excited bands to the zero-energy band.

In Fig.\ \ref{fig_ribbon_band_vs_slide}(c), we present the squared wave function of the flat band states at $\bar{K}$ of a bulk TBG (not of the ribbon) 
at $\lambda = 0, 0.2, \cdots, 1$. Here the orange dots represent AA spots, and the red and blue lines represent the boundary lines for the ribbon.
The wave amplitude is concentrated on the AA spots, which is a property of the flat band states \cite{PhysRevB.81.165105, PhysRevB.82.121407,PhysRevB.83.045425,PhysRevB.86.125413}.
In increasing $\lambda$, the bright spots on AA region shift upward to follow the moir\'{e} pattern movement.
The emergence of the edge states correlates with the relative position of the AA spots to the boundary lines.\cite{liu2019pseudo}
By comparing Figs.\ \ref{fig_ribbon_band_vs_slide}(a) and \ref{fig_ribbon_band_vs_slide}(c), we notice that
two edge states of the top boundary (red curves) branch out from the zero-energy flat band right when the AA spots cross the boundary to the outside,
and similarly, the two edge states of the lower boundary (blue curves) are absorbed into the flat band when the AA spots enter the ribbon from the lower boundary.

\begin{figure}
  \begin{center}
    \leavevmode\includegraphics[width=0.95 \hsize]{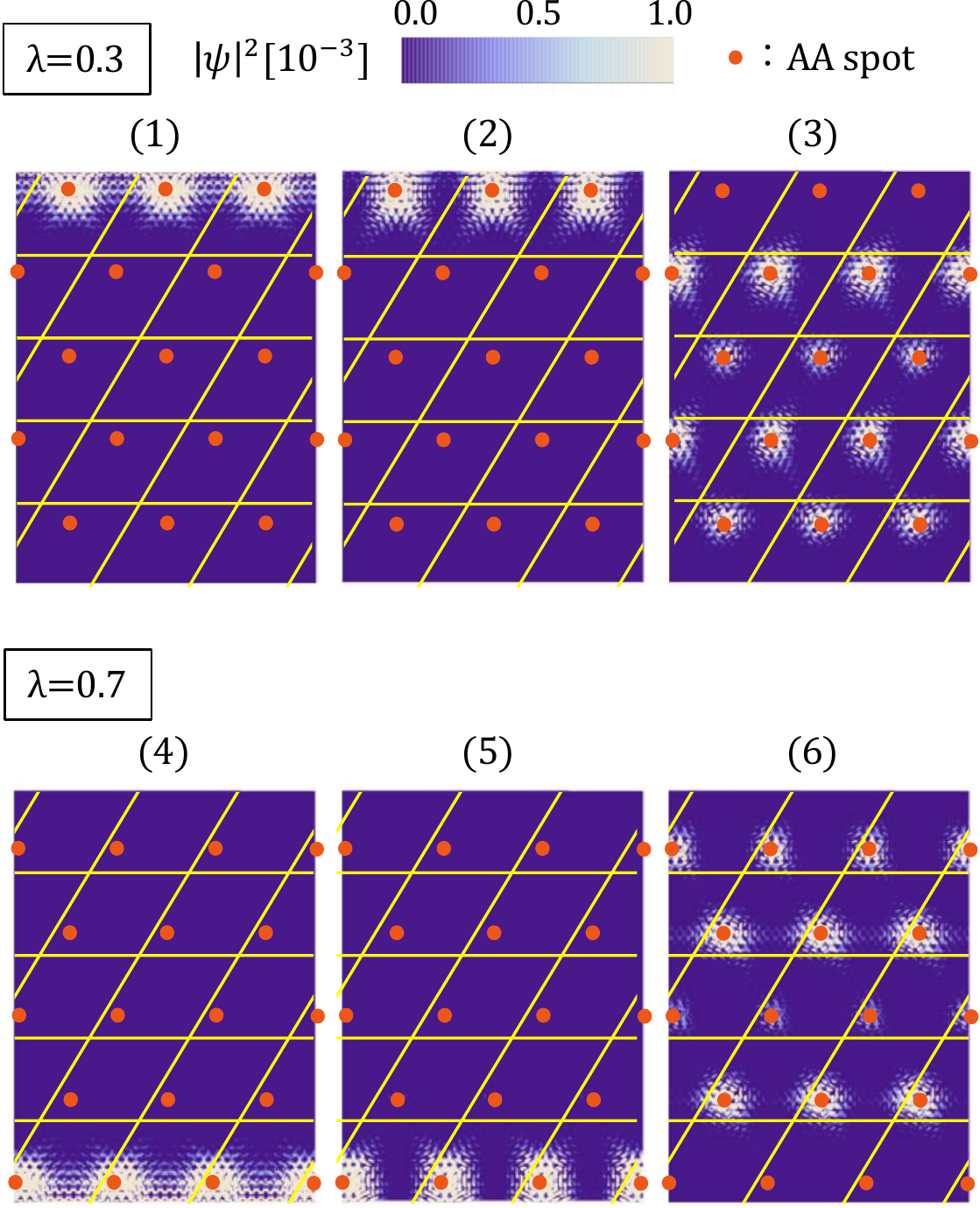}
    \caption{
    Real-space map of low-energy eigenstates of the TBG ribbon.
(1), (2) and (3) at $\lambda=0.3$, and (4), (5) and (6) at $\lambda=0.7$, which are labelled in  Fig.\ \ref{fig_ribbon_band_vs_slide}(a).
  }
    \label{fig_edge_wavefunction}
  \end{center}
  \end{figure}

Figure \ref{fig_edge_wavefunction} illustrates actual eigenstates of the TBG ribbon,
(1), (2) and (3) at $\lambda=0.3$, and (4), (5) and (6) at $\lambda=0.7$, which are labelled in  Fig.\ \ref{fig_ribbon_band_vs_slide}(a).
We see that the in-gap states (1) and (2) are actually localized on the upper edge, and (4) and (5) are on the lower edge,
while the bulk states (3) and (6) extend over the middle region.

\subsection{Bulk edge correspondence}

The number of edge states branching out or being absorbed per a sliding cycle ($0\leq \lambda \leq 1$)
exactly coincides with the sliding Chern number, which is a topological invariant defined for the Bloch bands of TBG  \cite{PhysRevB.101.041112,zhang2020topological,su2020topological}.
Let us consider an infinite TBG and assume the Fermi energy lies in an energy gap.
When we adiabatically slide the layer $l$ by its own lattice period $\mathbf{a}_{i}^{(l)}$ with the other layer fixed, 
the change of electronic polarization is written as \cite{PhysRevB.101.041112}
\begin{align}
\Delta \mathbf{P} =& C_{i1}^{(l)}\mathbf{L}^M_{1}+ C_{i2}^{(l)}\mathbf{L}^M_{2},
\\
C_{ij}^{(l)}=&
\sum_{n={\rm occupied}}
 \int_0^1 d k_j \int_{0}^{1} d\lambda_i 
 \nonumber\\
 &\quad
i  \left[
\Bigl\langle \frac{\partial u}{\partial \lambda_i} \Bigl| \frac{\partial u }{\partial k_j} \Bigr\rangle
-
\Bigl\langle \frac{\partial u}{\partial k_j} \Bigl| \frac{\partial u }{\partial \lambda_i} \Bigr\rangle
\right].
\end{align}
Here $u = u_{n\Vec{k}}(\lambda_1,\lambda_2)$ is the eigenstate of the $n$-th band at Bloch wave number $\Vec{k}$,
in the TBG Hamiltonian with the layer $l$ shifted by  $\lambda_1 \mathbf{a}_{1}^{(l)}+ \lambda_2 \mathbf{a}_{2}^{(l)}$.
The $k_j$ is a dimensionless wave component defined by $\Vec{k} = k_1 \Vec{G}^M_1+k_2 \Vec{G}^M_2$,
where $\mathbf{G}^M_{j}$ is the moir\'{e} reciprocal lattice vector satisfying $\mathbf{G}^M_{i}\cdot\mathbf{L}^M_{j} = 2\pi \delta_{ij}$.
$\Delta \mathbf{P}$ represents the shift of the polarization per a single moir\'{e} unit cell area during a sliding cycle .
%The sliding Chern number  $C_{ij}^{(l)}$ represents the number of electrons passed through the unit-cell side perpendicular to $\Vec{G}_{j}$
% (i.e., the cross section spanned by $\Vec{L}_2$ for $j=1$, and $\Vec{L}_1$ for $j=2$),
%during an adiabatic sliding of the layer $l$ by $\mathbf{a}_{i}^{(l)}$.

When the Fermi energy is in the gap just above the flat band, 
the sliding Chern numbers per spin are calculated as $C_{11}^{(1)}=C_{22}^{(1)}=2$, $C_{11}^{(2)}=C_{22}^{(2)}= -2$
and otherwise 0. \cite{PhysRevB.101.041112}
If we slide the layer $2$ by $\Delta\Vec{x}^{(2)} = (1/2)\mathbf{a}_{1}^{(2)}-\mathbf{a}_{2}^{(2)}$ as considered for the ribbon, 
the polarization shift per spin becomes
\begin{align}
\Delta \mathbf{P} &= 
\frac{1}{2} (C_{11}^{(2)}\mathbf{L}^M_{1}+ C_{12}^{(2)}\mathbf{L}^M_{2}) -
 (C_{21}^{(2)}\mathbf{L}^M_{1}+ C_{22}^{(2)}\mathbf{L}^M_{2}) 
\nonumber\\
& = 2 \left( -\frac{1}{2} \mathbf{L}_{1}^{M}+ \mathbf{L}_{2}^{M} \right) = 2 \Delta \Vec{X},
\end{align}
where  $\Delta \Vec{X}$ is the shift of the moir\'{e} pattern argued in the previous section.
This means that two electrons (per spin) pass through every unit-cell boundary perpendicular to $x$,
and it perfectly corresponds the number of the edge states branching out from (or being absorbed to) the flat band
per a sliding cycle, as argued in  Fig.\ \ref{fig_ribbon_band_vs_slide}(a). 

Figure \ref{fig_ribbon_band_vs_slide}(a) also shows that,  in the second gap around 0.8 eV,
four bands of the upper edge states go up, and four bands of the lower edge states go down
 in a sliding process from $\lambda=0$ to 1.
Correspondingly, the sliding Chern numbers (per spin) in the gap are $C_{11}^{(1)}=C_{22}^{(1)}=4$, $C_{11}^{(2)}=C_{22}^{(2)}= -4$ and otherwise 0,
giving  $\Delta \mathbf{P} = 4 \Delta \Vec{X}$.

This is analogous to the bulk-edge correspondence for the quantum Hall effect, \cite{PhysRevLett.71.3697,PhysRevB.48.11851}
where a similar transfer of the edge states between energy bands is observed against
a change of the Bloch wave number along the ribbon.
In the present case, it occurs as a function as a sliding parameter $\lambda$.
Unlike the quantum Hall case, the edge states corresponding to the sliding Chern number
do not fill out the whole energy gap in a single static physical system with $\lambda$ fixed,
but it fills out the gap when $\lambda$ is changed by a cycle.
We note that the bulk-edge correspondence for the topological pumping was first 
proposed for a one-dimensional lattice with time-dependent moving potential. \cite{hatsugai2016bulk}

%%%%%%%%%%%%%
\section{Conclusion}
\label{sec_concl}
We studied the edge states of TBG and the topological correspondence to relative interlayer sliding. 
By calculating the eigenfunctions of the edge-terminated TBG ribbon as a function of the sliding distance,
we demonstrated that moir\'{e} edge states are transferred across the band gap during the interlayer sliding, and
the number of edge states pumped in the sliding cycle coincides with the sliding Chern number of the band gap. 
The relationship can be viewed as a manifestation of the bulk-edge correspondence in moir\'{e} bilayer systems,
where nonzero sliding Chern number is always associated with the emergence of the moir\'{e} edge states.

\section*{Acknowledgments}
This work is supported by JSPS KAKENHI Grant Number JP20H01840 and JP20H00127, Japan,
and by JST CREST Grant Number JPMJCR20T3, Japan.

%%%%%%%%%%%%%%%%%%%%%%%%%%%
%%%%%%%%%%%%%%%%%%%%%%%%%%%
%%%%%%%%%%%%%%%%%%%%%%%%%%%

\bibliography{moire_pump_edge.bib}

\begin{thebibliography}{48}%
\makeatletter
\providecommand \@ifxundefined [1]{%
 \@ifx{#1\undefined}
}%
\providecommand \@ifnum [1]{%
 \ifnum #1\expandafter \@firstoftwo
 \else \expandafter \@secondoftwo
 \fi
}%
\providecommand \@ifx [1]{%
 \ifx #1\expandafter \@firstoftwo
 \else \expandafter \@secondoftwo
 \fi
}%
\providecommand \natexlab [1]{#1}%
\providecommand \enquote  [1]{``#1''}%
\providecommand \bibnamefont  [1]{#1}%
\providecommand \bibfnamefont [1]{#1}%
\providecommand \citenamefont [1]{#1}%
\providecommand \href@noop [0]{\@secondoftwo}%
\providecommand \href [0]{\begingroup \@sanitize@url \@href}%
\providecommand \@href[1]{\@@startlink{#1}\@@href}%
\providecommand \@@href[1]{\endgroup#1\@@endlink}%
\providecommand \@sanitize@url [0]{\catcode `\\12\catcode `\$12\catcode
  `\&12\catcode `\#12\catcode `\^12\catcode `\_12\catcode `\%12\relax}%
\providecommand \@@startlink[1]{}%
\providecommand \@@endlink[0]{}%
\providecommand \url  [0]{\begingroup\@sanitize@url \@url }%
\providecommand \@url [1]{\endgroup\@href {#1}{\urlprefix }}%
\providecommand \urlprefix  [0]{URL }%
\providecommand \Eprint [0]{\href }%
\providecommand \doibase [0]{http://dx.doi.org/}%
\providecommand \selectlanguage [0]{\@gobble}%
\providecommand \bibinfo  [0]{\@secondoftwo}%
\providecommand \bibfield  [0]{\@secondoftwo}%
\providecommand \translation [1]{[#1]}%
\providecommand \BibitemOpen [0]{}%
\providecommand \bibitemStop [0]{}%
\providecommand \bibitemNoStop [0]{.\EOS\space}%
\providecommand \EOS [0]{\spacefactor3000\relax}%
\providecommand \BibitemShut  [1]{\csname bibitem#1\endcsname}%
\let\auto@bib@innerbib\@empty
%</preamble>
\bibitem [{\citenamefont {Thouless}\ \emph {et~al.}(1982)\citenamefont
  {Thouless}, \citenamefont {Kohmoto}, \citenamefont {Nightingale},\ and\
  \citenamefont {den Nijs}}]{PhysRevLett.49.405}%
  \BibitemOpen
  \bibfield  {author} {\bibinfo {author} {\bibfnamefont {D.~J.}\ \bibnamefont
  {Thouless}}, \bibinfo {author} {\bibfnamefont {M.}~\bibnamefont {Kohmoto}},
  \bibinfo {author} {\bibfnamefont {M.~P.}\ \bibnamefont {Nightingale}}, \ and\
  \bibinfo {author} {\bibfnamefont {M.}~\bibnamefont {den Nijs}},\ }\href
  {\doibase 10.1103/PhysRevLett.49.405} {\bibfield  {journal} {\bibinfo
  {journal} {Phys. Rev. Lett.}\ }\textbf {\bibinfo {volume} {49}},\ \bibinfo
  {pages} {405} (\bibinfo {year} {1982})}\BibitemShut {NoStop}%
\bibitem [{\citenamefont {Kane}\ and\ \citenamefont {Mele}(2005)}]{kane2005z}%
  \BibitemOpen
  \bibfield  {author} {\bibinfo {author} {\bibfnamefont {C.~L.}\ \bibnamefont
  {Kane}}\ and\ \bibinfo {author} {\bibfnamefont {E.~J.}\ \bibnamefont
  {Mele}},\ }\href@noop {} {\bibfield  {journal} {\bibinfo  {journal} {Physical
  review letters}\ }\textbf {\bibinfo {volume} {95}},\ \bibinfo {pages}
  {146802} (\bibinfo {year} {2005})}\BibitemShut {NoStop}%
\bibitem [{\citenamefont {Qi}\ \emph {et~al.}(2008)\citenamefont {Qi},
  \citenamefont {Hughes},\ and\ \citenamefont {Zhang}}]{qi2008topological}%
  \BibitemOpen
  \bibfield  {author} {\bibinfo {author} {\bibfnamefont {X.-L.}\ \bibnamefont
  {Qi}}, \bibinfo {author} {\bibfnamefont {T.~L.}\ \bibnamefont {Hughes}}, \
  and\ \bibinfo {author} {\bibfnamefont {S.-C.}\ \bibnamefont {Zhang}},\
  }\href@noop {} {\bibfield  {journal} {\bibinfo  {journal} {Physical Review
  B}\ }\textbf {\bibinfo {volume} {78}},\ \bibinfo {pages} {195424} (\bibinfo
  {year} {2008})}\BibitemShut {NoStop}%
\bibitem [{\citenamefont {Hatsugai}(1993{\natexlab{a}})}]{PhysRevLett.71.3697}%
  \BibitemOpen
  \bibfield  {author} {\bibinfo {author} {\bibfnamefont {Y.}~\bibnamefont
  {Hatsugai}},\ }\href {\doibase 10.1103/PhysRevLett.71.3697} {\bibfield
  {journal} {\bibinfo  {journal} {Phys. Rev. Lett.}\ }\textbf {\bibinfo
  {volume} {71}},\ \bibinfo {pages} {3697} (\bibinfo {year}
  {1993}{\natexlab{a}})}\BibitemShut {NoStop}%
\bibitem [{\citenamefont {Hatsugai}(1993{\natexlab{b}})}]{PhysRevB.48.11851}%
  \BibitemOpen
  \bibfield  {author} {\bibinfo {author} {\bibfnamefont {Y.}~\bibnamefont
  {Hatsugai}},\ }\href {\doibase 10.1103/PhysRevB.48.11851} {\bibfield
  {journal} {\bibinfo  {journal} {Phys. Rev. B}\ }\textbf {\bibinfo {volume}
  {48}},\ \bibinfo {pages} {11851} (\bibinfo {year}
  {1993}{\natexlab{b}})}\BibitemShut {NoStop}%
\bibitem [{\citenamefont {Kohmoto}(1985)}]{KOHMOTO1985343}%
  \BibitemOpen
  \bibfield  {author} {\bibinfo {author} {\bibfnamefont {M.}~\bibnamefont
  {Kohmoto}},\ }\href {\doibase https://doi.org/10.1016/0003-4916(85)90148-4}
  {\bibfield  {journal} {\bibinfo  {journal} {Annals of Physics}\ }\textbf
  {\bibinfo {volume} {160}},\ \bibinfo {pages} {343 } (\bibinfo {year}
  {1985})}\BibitemShut {NoStop}%
\bibitem [{\citenamefont {Hasan}\ and\ \citenamefont
  {Kane}(2010)}]{hasan2010colloquium}%
  \BibitemOpen
  \bibfield  {author} {\bibinfo {author} {\bibfnamefont {M.~Z.}\ \bibnamefont
  {Hasan}}\ and\ \bibinfo {author} {\bibfnamefont {C.~L.}\ \bibnamefont
  {Kane}},\ }\href@noop {} {\bibfield  {journal} {\bibinfo  {journal} {Reviews
  of modern physics}\ }\textbf {\bibinfo {volume} {82}},\ \bibinfo {pages}
  {3045} (\bibinfo {year} {2010})}\BibitemShut {NoStop}%
\bibitem [{\citenamefont {Qi}\ and\ \citenamefont
  {Zhang}(2011)}]{qi2011topological}%
  \BibitemOpen
  \bibfield  {author} {\bibinfo {author} {\bibfnamefont {X.-L.}\ \bibnamefont
  {Qi}}\ and\ \bibinfo {author} {\bibfnamefont {S.-C.}\ \bibnamefont {Zhang}},\
  }\href@noop {} {\bibfield  {journal} {\bibinfo  {journal} {Reviews of Modern
  Physics}\ }\textbf {\bibinfo {volume} {83}},\ \bibinfo {pages} {1057}
  (\bibinfo {year} {2011})}\BibitemShut {NoStop}%
\bibitem [{\citenamefont {Kane}\ and\ \citenamefont
  {Lubensky}(2014)}]{kane2014topological}%
  \BibitemOpen
  \bibfield  {author} {\bibinfo {author} {\bibfnamefont {C.}~\bibnamefont
  {Kane}}\ and\ \bibinfo {author} {\bibfnamefont {T.}~\bibnamefont
  {Lubensky}},\ }\href@noop {} {\bibfield  {journal} {\bibinfo  {journal}
  {Nature Physics}\ }\textbf {\bibinfo {volume} {10}},\ \bibinfo {pages} {39}
  (\bibinfo {year} {2014})}\BibitemShut {NoStop}%
\bibitem [{\citenamefont {Kariyado}\ and\ \citenamefont
  {Hatsugai}(2015)}]{kariyado2015manipulation}%
  \BibitemOpen
  \bibfield  {author} {\bibinfo {author} {\bibfnamefont {T.}~\bibnamefont
  {Kariyado}}\ and\ \bibinfo {author} {\bibfnamefont {Y.}~\bibnamefont
  {Hatsugai}},\ }\href@noop {} {\bibfield  {journal} {\bibinfo  {journal}
  {Scientific reports}\ }\textbf {\bibinfo {volume} {5}},\ \bibinfo {pages}
  {18107} (\bibinfo {year} {2015})}\BibitemShut {NoStop}%
\bibitem [{\citenamefont {S{\"u}sstrunk}\ and\ \citenamefont
  {Huber}(2016)}]{sstrunkE4767}%
  \BibitemOpen
  \bibfield  {author} {\bibinfo {author} {\bibfnamefont {R.}~\bibnamefont
  {S{\"u}sstrunk}}\ and\ \bibinfo {author} {\bibfnamefont {S.~D.}\ \bibnamefont
  {Huber}},\ }\href {\doibase 10.1073/pnas.1605462113} {\bibfield  {journal}
  {\bibinfo  {journal} {Proceedings of the National Academy of Sciences}\
  }\textbf {\bibinfo {volume} {113}},\ \bibinfo {pages} {E4767} (\bibinfo
  {year} {2016})}\BibitemShut {NoStop}%
\bibitem [{\citenamefont {Ozawa}\ \emph {et~al.}(2019)\citenamefont {Ozawa},
  \citenamefont {Price}, \citenamefont {Amo}, \citenamefont {Goldman},
  \citenamefont {Hafezi}, \citenamefont {Lu}, \citenamefont {Rechtsman},
  \citenamefont {Schuster}, \citenamefont {Simon}, \citenamefont {Zilberberg},\
  and\ \citenamefont {Carusotto}}]{RevModPhys.91.015006}%
  \BibitemOpen
  \bibfield  {author} {\bibinfo {author} {\bibfnamefont {T.}~\bibnamefont
  {Ozawa}}, \bibinfo {author} {\bibfnamefont {H.~M.}\ \bibnamefont {Price}},
  \bibinfo {author} {\bibfnamefont {A.}~\bibnamefont {Amo}}, \bibinfo {author}
  {\bibfnamefont {N.}~\bibnamefont {Goldman}}, \bibinfo {author} {\bibfnamefont
  {M.}~\bibnamefont {Hafezi}}, \bibinfo {author} {\bibfnamefont
  {L.}~\bibnamefont {Lu}}, \bibinfo {author} {\bibfnamefont {M.~C.}\
  \bibnamefont {Rechtsman}}, \bibinfo {author} {\bibfnamefont {D.}~\bibnamefont
  {Schuster}}, \bibinfo {author} {\bibfnamefont {J.}~\bibnamefont {Simon}},
  \bibinfo {author} {\bibfnamefont {O.}~\bibnamefont {Zilberberg}}, \ and\
  \bibinfo {author} {\bibfnamefont {I.}~\bibnamefont {Carusotto}},\ }\href
  {\doibase 10.1103/RevModPhys.91.015006} {\bibfield  {journal} {\bibinfo
  {journal} {Rev. Mod. Phys.}\ }\textbf {\bibinfo {volume} {91}},\ \bibinfo
  {pages} {015006} (\bibinfo {year} {2019})}\BibitemShut {NoStop}%
\bibitem [{\citenamefont {Lu}\ \emph {et~al.}(2014)\citenamefont {Lu},
  \citenamefont {Joannopoulos},\ and\ \citenamefont
  {Solja{\v{c}}i{\'c}}}]{lu2014topological}%
  \BibitemOpen
  \bibfield  {author} {\bibinfo {author} {\bibfnamefont {L.}~\bibnamefont
  {Lu}}, \bibinfo {author} {\bibfnamefont {J.~D.}\ \bibnamefont
  {Joannopoulos}}, \ and\ \bibinfo {author} {\bibfnamefont {M.}~\bibnamefont
  {Solja{\v{c}}i{\'c}}},\ }\href@noop {} {\bibfield  {journal} {\bibinfo
  {journal} {Nature photonics}\ }\textbf {\bibinfo {volume} {8}},\ \bibinfo
  {pages} {821} (\bibinfo {year} {2014})}\BibitemShut {NoStop}%
\bibitem [{\citenamefont {Lopes~dos Santos}\ \emph {et~al.}(2007)\citenamefont
  {Lopes~dos Santos}, \citenamefont {Peres},\ and\ \citenamefont
  {Castro~Neto}}]{PhysRevLett.99.256802}%
  \BibitemOpen
  \bibfield  {author} {\bibinfo {author} {\bibfnamefont {J.~M.~B.}\
  \bibnamefont {Lopes~dos Santos}}, \bibinfo {author} {\bibfnamefont
  {N.~M.~R.}\ \bibnamefont {Peres}}, \ and\ \bibinfo {author} {\bibfnamefont
  {A.~H.}\ \bibnamefont {Castro~Neto}},\ }\href {\doibase
  10.1103/PhysRevLett.99.256802} {\bibfield  {journal} {\bibinfo  {journal}
  {Phys. Rev. Lett.}\ }\textbf {\bibinfo {volume} {99}},\ \bibinfo {pages}
  {256802} (\bibinfo {year} {2007})}\BibitemShut {NoStop}%
\bibitem [{\citenamefont {Mele}(2010)}]{PhysRevB.81.161405}%
  \BibitemOpen
  \bibfield  {author} {\bibinfo {author} {\bibfnamefont {E.~J.}\ \bibnamefont
  {Mele}},\ }\href {\doibase 10.1103/PhysRevB.81.161405} {\bibfield  {journal}
  {\bibinfo  {journal} {Phys. Rev. B}\ }\textbf {\bibinfo {volume} {81}},\
  \bibinfo {pages} {161405} (\bibinfo {year} {2010})}\BibitemShut {NoStop}%
\bibitem [{\citenamefont {Trambly~de Laissardi{\`e}re}\ \emph
  {et~al.}(2010)\citenamefont {Trambly~de Laissardi{\`e}re}, \citenamefont
  {Mayou},\ and\ \citenamefont {Magaud}}]{trambly2010localization}%
  \BibitemOpen
  \bibfield  {author} {\bibinfo {author} {\bibfnamefont {G.}~\bibnamefont
  {Trambly~de Laissardi{\`e}re}}, \bibinfo {author} {\bibfnamefont
  {D.}~\bibnamefont {Mayou}}, \ and\ \bibinfo {author} {\bibfnamefont
  {L.}~\bibnamefont {Magaud}},\ }\href@noop {} {\bibfield  {journal} {\bibinfo
  {journal} {Nano letters}\ }\textbf {\bibinfo {volume} {10}},\ \bibinfo
  {pages} {804} (\bibinfo {year} {2010})}\BibitemShut {NoStop}%
\bibitem [{\citenamefont {Shallcross}\ \emph {et~al.}(2010)\citenamefont
  {Shallcross}, \citenamefont {Sharma}, \citenamefont {Kandelaki},\ and\
  \citenamefont {Pankratov}}]{PhysRevB.81.165105}%
  \BibitemOpen
  \bibfield  {author} {\bibinfo {author} {\bibfnamefont {S.}~\bibnamefont
  {Shallcross}}, \bibinfo {author} {\bibfnamefont {S.}~\bibnamefont {Sharma}},
  \bibinfo {author} {\bibfnamefont {E.}~\bibnamefont {Kandelaki}}, \ and\
  \bibinfo {author} {\bibfnamefont {O.~A.}\ \bibnamefont {Pankratov}},\ }\href
  {\doibase 10.1103/PhysRevB.81.165105} {\bibfield  {journal} {\bibinfo
  {journal} {Phys. Rev. B}\ }\textbf {\bibinfo {volume} {81}},\ \bibinfo
  {pages} {165105} (\bibinfo {year} {2010})}\BibitemShut {NoStop}%
\bibitem [{\citenamefont {Su\'arez~Morell}\ \emph {et~al.}(2010)\citenamefont
  {Su\'arez~Morell}, \citenamefont {Correa}, \citenamefont {Vargas},
  \citenamefont {Pacheco},\ and\ \citenamefont
  {Barticevic}}]{PhysRevB.82.121407}%
  \BibitemOpen
  \bibfield  {author} {\bibinfo {author} {\bibfnamefont {E.}~\bibnamefont
  {Su\'arez~Morell}}, \bibinfo {author} {\bibfnamefont {J.~D.}\ \bibnamefont
  {Correa}}, \bibinfo {author} {\bibfnamefont {P.}~\bibnamefont {Vargas}},
  \bibinfo {author} {\bibfnamefont {M.}~\bibnamefont {Pacheco}}, \ and\
  \bibinfo {author} {\bibfnamefont {Z.}~\bibnamefont {Barticevic}},\ }\href
  {\doibase 10.1103/PhysRevB.82.121407} {\bibfield  {journal} {\bibinfo
  {journal} {Phys. Rev. B}\ }\textbf {\bibinfo {volume} {82}},\ \bibinfo
  {pages} {121407} (\bibinfo {year} {2010})}\BibitemShut {NoStop}%
\bibitem [{\citenamefont {Bistritzer}\ and\ \citenamefont
  {MacDonald}(2011)}]{bistritzer2011moire}%
  \BibitemOpen
  \bibfield  {author} {\bibinfo {author} {\bibfnamefont {R.}~\bibnamefont
  {Bistritzer}}\ and\ \bibinfo {author} {\bibfnamefont {A.~H.}\ \bibnamefont
  {MacDonald}},\ }\href@noop {} {\bibfield  {journal} {\bibinfo  {journal}
  {Proceedings of the National Academy of Sciences}\ }\textbf {\bibinfo
  {volume} {108}},\ \bibinfo {pages} {12233} (\bibinfo {year}
  {2011})}\BibitemShut {NoStop}%
\bibitem [{\citenamefont {Kindermann}\ and\ \citenamefont
  {First}(2011)}]{PhysRevB.83.045425}%
  \BibitemOpen
  \bibfield  {author} {\bibinfo {author} {\bibfnamefont {M.}~\bibnamefont
  {Kindermann}}\ and\ \bibinfo {author} {\bibfnamefont {P.~N.}\ \bibnamefont
  {First}},\ }\href {\doibase 10.1103/PhysRevB.83.045425} {\bibfield  {journal}
  {\bibinfo  {journal} {Phys. Rev. B}\ }\textbf {\bibinfo {volume} {83}},\
  \bibinfo {pages} {045425} (\bibinfo {year} {2011})}\BibitemShut {NoStop}%
\bibitem [{\citenamefont {Xian}\ \emph {et~al.}(2011)\citenamefont {Xian},
  \citenamefont {Barraza-Lopez},\ and\ \citenamefont
  {Chou}}]{PhysRevB.84.075425}%
  \BibitemOpen
  \bibfield  {author} {\bibinfo {author} {\bibfnamefont {L.}~\bibnamefont
  {Xian}}, \bibinfo {author} {\bibfnamefont {S.}~\bibnamefont {Barraza-Lopez}},
  \ and\ \bibinfo {author} {\bibfnamefont {M.~Y.}\ \bibnamefont {Chou}},\
  }\href {\doibase 10.1103/PhysRevB.84.075425} {\bibfield  {journal} {\bibinfo
  {journal} {Phys. Rev. B}\ }\textbf {\bibinfo {volume} {84}},\ \bibinfo
  {pages} {075425} (\bibinfo {year} {2011})}\BibitemShut {NoStop}%
\bibitem [{\citenamefont {Dos~Santos}\ \emph {et~al.}(2012)\citenamefont
  {Dos~Santos}, \citenamefont {Peres},\ and\ \citenamefont
  {Neto}}]{dos2012continuum}%
  \BibitemOpen
  \bibfield  {author} {\bibinfo {author} {\bibfnamefont {J.~L.}\ \bibnamefont
  {Dos~Santos}}, \bibinfo {author} {\bibfnamefont {N.}~\bibnamefont {Peres}}, \
  and\ \bibinfo {author} {\bibfnamefont {A.~C.}\ \bibnamefont {Neto}},\
  }\href@noop {} {\bibfield  {journal} {\bibinfo  {journal} {Physical Review
  B}\ }\textbf {\bibinfo {volume} {86}},\ \bibinfo {pages} {155449} (\bibinfo
  {year} {2012})}\BibitemShut {NoStop}%
\bibitem [{\citenamefont {Moon}\ and\ \citenamefont
  {Koshino}(2012)}]{PhysRevB.85.195458}%
  \BibitemOpen
  \bibfield  {author} {\bibinfo {author} {\bibfnamefont {P.}~\bibnamefont
  {Moon}}\ and\ \bibinfo {author} {\bibfnamefont {M.}~\bibnamefont {Koshino}},\
  }\href {\doibase 10.1103/PhysRevB.85.195458} {\bibfield  {journal} {\bibinfo
  {journal} {Phys. Rev. B}\ }\textbf {\bibinfo {volume} {85}},\ \bibinfo
  {pages} {195458} (\bibinfo {year} {2012})}\BibitemShut {NoStop}%
\bibitem [{\citenamefont {Trambly~de Laissardi\`ere}\ \emph
  {et~al.}(2012)\citenamefont {Trambly~de Laissardi\`ere}, \citenamefont
  {Mayou},\ and\ \citenamefont {Magaud}}]{PhysRevB.86.125413}%
  \BibitemOpen
  \bibfield  {author} {\bibinfo {author} {\bibfnamefont {G.}~\bibnamefont
  {Trambly~de Laissardi\`ere}}, \bibinfo {author} {\bibfnamefont
  {D.}~\bibnamefont {Mayou}}, \ and\ \bibinfo {author} {\bibfnamefont
  {L.}~\bibnamefont {Magaud}},\ }\href {\doibase 10.1103/PhysRevB.86.125413}
  {\bibfield  {journal} {\bibinfo  {journal} {Phys. Rev. B}\ }\textbf {\bibinfo
  {volume} {86}},\ \bibinfo {pages} {125413} (\bibinfo {year}
  {2012})}\BibitemShut {NoStop}%
\bibitem [{\citenamefont {Moon}\ and\ \citenamefont
  {Koshino}(2013)}]{PhysRevB.87.205404}%
  \BibitemOpen
  \bibfield  {author} {\bibinfo {author} {\bibfnamefont {P.}~\bibnamefont
  {Moon}}\ and\ \bibinfo {author} {\bibfnamefont {M.}~\bibnamefont {Koshino}},\
  }\href {\doibase 10.1103/PhysRevB.87.205404} {\bibfield  {journal} {\bibinfo
  {journal} {Phys. Rev. B}\ }\textbf {\bibinfo {volume} {87}},\ \bibinfo
  {pages} {205404} (\bibinfo {year} {2013})}\BibitemShut {NoStop}%
\bibitem [{\citenamefont {Cao}\ \emph {et~al.}(2016)\citenamefont {Cao},
  \citenamefont {Luo}, \citenamefont {Fatemi}, \citenamefont {Fang},
  \citenamefont {Sanchez-Yamagishi}, \citenamefont {Watanabe}, \citenamefont
  {Taniguchi}, \citenamefont {Kaxiras},\ and\ \citenamefont
  {Jarillo-Herrero}}]{PhysRevLett.117.116804}%
  \BibitemOpen
  \bibfield  {author} {\bibinfo {author} {\bibfnamefont {Y.}~\bibnamefont
  {Cao}}, \bibinfo {author} {\bibfnamefont {J.~Y.}\ \bibnamefont {Luo}},
  \bibinfo {author} {\bibfnamefont {V.}~\bibnamefont {Fatemi}}, \bibinfo
  {author} {\bibfnamefont {S.}~\bibnamefont {Fang}}, \bibinfo {author}
  {\bibfnamefont {J.~D.}\ \bibnamefont {Sanchez-Yamagishi}}, \bibinfo {author}
  {\bibfnamefont {K.}~\bibnamefont {Watanabe}}, \bibinfo {author}
  {\bibfnamefont {T.}~\bibnamefont {Taniguchi}}, \bibinfo {author}
  {\bibfnamefont {E.}~\bibnamefont {Kaxiras}}, \ and\ \bibinfo {author}
  {\bibfnamefont {P.}~\bibnamefont {Jarillo-Herrero}},\ }\href {\doibase
  10.1103/PhysRevLett.117.116804} {\bibfield  {journal} {\bibinfo  {journal}
  {Phys. Rev. Lett.}\ }\textbf {\bibinfo {volume} {117}},\ \bibinfo {pages}
  {116804} (\bibinfo {year} {2016})}\BibitemShut {NoStop}%
\bibitem [{\citenamefont {Cao}\ \emph {et~al.}(2018{\natexlab{a}})\citenamefont
  {Cao}, \citenamefont {Fatemi}, \citenamefont {Fang}, \citenamefont
  {Watanabe}, \citenamefont {Taniguchi}, \citenamefont {Kaxiras},\ and\
  \citenamefont {Jarillo-Herrero}}]{cao2018unconventional}%
  \BibitemOpen
  \bibfield  {author} {\bibinfo {author} {\bibfnamefont {Y.}~\bibnamefont
  {Cao}}, \bibinfo {author} {\bibfnamefont {V.}~\bibnamefont {Fatemi}},
  \bibinfo {author} {\bibfnamefont {S.}~\bibnamefont {Fang}}, \bibinfo {author}
  {\bibfnamefont {K.}~\bibnamefont {Watanabe}}, \bibinfo {author}
  {\bibfnamefont {T.}~\bibnamefont {Taniguchi}}, \bibinfo {author}
  {\bibfnamefont {E.}~\bibnamefont {Kaxiras}}, \ and\ \bibinfo {author}
  {\bibfnamefont {P.}~\bibnamefont {Jarillo-Herrero}},\ }\href@noop {}
  {\bibfield  {journal} {\bibinfo  {journal} {Nature}\ }\textbf {\bibinfo
  {volume} {556}},\ \bibinfo {pages} {43} (\bibinfo {year}
  {2018}{\natexlab{a}})}\BibitemShut {NoStop}%
\bibitem [{\citenamefont {Landgraf}\ \emph {et~al.}(2013)\citenamefont
  {Landgraf}, \citenamefont {Shallcross}, \citenamefont {T\"urschmann},
  \citenamefont {Weckbecker},\ and\ \citenamefont
  {Pankratov}}]{PhysRevB.87.075433}%
  \BibitemOpen
  \bibfield  {author} {\bibinfo {author} {\bibfnamefont {W.}~\bibnamefont
  {Landgraf}}, \bibinfo {author} {\bibfnamefont {S.}~\bibnamefont
  {Shallcross}}, \bibinfo {author} {\bibfnamefont {K.}~\bibnamefont
  {T\"urschmann}}, \bibinfo {author} {\bibfnamefont {D.}~\bibnamefont
  {Weckbecker}}, \ and\ \bibinfo {author} {\bibfnamefont {O.}~\bibnamefont
  {Pankratov}},\ }\href {\doibase 10.1103/PhysRevB.87.075433} {\bibfield
  {journal} {\bibinfo  {journal} {Phys. Rev. B}\ }\textbf {\bibinfo {volume}
  {87}},\ \bibinfo {pages} {075433} (\bibinfo {year} {2013})}\BibitemShut
  {NoStop}%
\bibitem [{\citenamefont {Su\'arez~Morell}\ \emph {et~al.}(2014)\citenamefont
  {Su\'arez~Morell}, \citenamefont {Vergara}, \citenamefont {Pacheco},
  \citenamefont {Brey},\ and\ \citenamefont {Chico}}]{PhysRevB.89.205405}%
  \BibitemOpen
  \bibfield  {author} {\bibinfo {author} {\bibfnamefont {E.}~\bibnamefont
  {Su\'arez~Morell}}, \bibinfo {author} {\bibfnamefont {R.}~\bibnamefont
  {Vergara}}, \bibinfo {author} {\bibfnamefont {M.}~\bibnamefont {Pacheco}},
  \bibinfo {author} {\bibfnamefont {L.}~\bibnamefont {Brey}}, \ and\ \bibinfo
  {author} {\bibfnamefont {L.}~\bibnamefont {Chico}},\ }\href {\doibase
  10.1103/PhysRevB.89.205405} {\bibfield  {journal} {\bibinfo  {journal} {Phys.
  Rev. B}\ }\textbf {\bibinfo {volume} {89}},\ \bibinfo {pages} {205405}
  (\bibinfo {year} {2014})}\BibitemShut {NoStop}%
\bibitem [{\citenamefont {Morell}\ \emph {et~al.}(2015)\citenamefont {Morell},
  \citenamefont {Vargas}, \citenamefont {H\"aberle}, \citenamefont {Hevia},\
  and\ \citenamefont {Chico}}]{PhysRevB.91.035441}%
  \BibitemOpen
  \bibfield  {author} {\bibinfo {author} {\bibfnamefont {E.~S.}\ \bibnamefont
  {Morell}}, \bibinfo {author} {\bibfnamefont {P.}~\bibnamefont {Vargas}},
  \bibinfo {author} {\bibfnamefont {P.}~\bibnamefont {H\"aberle}}, \bibinfo
  {author} {\bibfnamefont {S.~A.}\ \bibnamefont {Hevia}}, \ and\ \bibinfo
  {author} {\bibfnamefont {L.}~\bibnamefont {Chico}},\ }\href {\doibase
  10.1103/PhysRevB.91.035441} {\bibfield  {journal} {\bibinfo  {journal} {Phys.
  Rev. B}\ }\textbf {\bibinfo {volume} {91}},\ \bibinfo {pages} {035441}
  (\bibinfo {year} {2015})}\BibitemShut {NoStop}%
\bibitem [{\citenamefont {Pelc}\ \emph {et~al.}(2015)\citenamefont {Pelc},
  \citenamefont {Morell}, \citenamefont {Brey},\ and\ \citenamefont
  {Chico}}]{pelc2015electronic}%
  \BibitemOpen
  \bibfield  {author} {\bibinfo {author} {\bibfnamefont {M.}~\bibnamefont
  {Pelc}}, \bibinfo {author} {\bibfnamefont {E.~S.}\ \bibnamefont {Morell}},
  \bibinfo {author} {\bibfnamefont {L.}~\bibnamefont {Brey}}, \ and\ \bibinfo
  {author} {\bibfnamefont {L.}~\bibnamefont {Chico}},\ }\href@noop {}
  {\bibfield  {journal} {\bibinfo  {journal} {J. Phys. Chem. C}\ }\textbf
  {\bibinfo {volume} {119}},\ \bibinfo {pages} {10076} (\bibinfo {year}
  {2015})}\BibitemShut {NoStop}%
\bibitem [{\citenamefont {Fleischmann}\ \emph {et~al.}(2018)\citenamefont
  {Fleischmann}, \citenamefont {Gupta}, \citenamefont {Weckbecker},
  \citenamefont {Landgraf}, \citenamefont {Pankratov}, \citenamefont {Meded},\
  and\ \citenamefont {Shallcross}}]{fleischmann2018moire}%
  \BibitemOpen
  \bibfield  {author} {\bibinfo {author} {\bibfnamefont {M.}~\bibnamefont
  {Fleischmann}}, \bibinfo {author} {\bibfnamefont {R.}~\bibnamefont {Gupta}},
  \bibinfo {author} {\bibfnamefont {D.}~\bibnamefont {Weckbecker}}, \bibinfo
  {author} {\bibfnamefont {W.}~\bibnamefont {Landgraf}}, \bibinfo {author}
  {\bibfnamefont {O.}~\bibnamefont {Pankratov}}, \bibinfo {author}
  {\bibfnamefont {V.}~\bibnamefont {Meded}}, \ and\ \bibinfo {author}
  {\bibfnamefont {S.}~\bibnamefont {Shallcross}},\ }\href@noop {} {\bibfield
  {journal} {\bibinfo  {journal} {Physical Review B}\ }\textbf {\bibinfo
  {volume} {97}},\ \bibinfo {pages} {205128} (\bibinfo {year}
  {2018})}\BibitemShut {NoStop}%
\bibitem [{\citenamefont {Liu}\ \emph {et~al.}(2019)\citenamefont {Liu},
  \citenamefont {Liu},\ and\ \citenamefont {Dai}}]{liu2019pseudo}%
  \BibitemOpen
  \bibfield  {author} {\bibinfo {author} {\bibfnamefont {J.}~\bibnamefont
  {Liu}}, \bibinfo {author} {\bibfnamefont {J.}~\bibnamefont {Liu}}, \ and\
  \bibinfo {author} {\bibfnamefont {X.}~\bibnamefont {Dai}},\ }\href@noop {}
  {\bibfield  {journal} {\bibinfo  {journal} {Physical Review B}\ }\textbf
  {\bibinfo {volume} {99}},\ \bibinfo {pages} {155415} (\bibinfo {year}
  {2019})}\BibitemShut {NoStop}%
\bibitem [{\citenamefont {Wakabayashi}\ \emph {et~al.}(1999)\citenamefont
  {Wakabayashi}, \citenamefont {Fujita}, \citenamefont {Ajiki},\ and\
  \citenamefont {Sigrist}}]{PhysRevB.59.8271}%
  \BibitemOpen
  \bibfield  {author} {\bibinfo {author} {\bibfnamefont {K.}~\bibnamefont
  {Wakabayashi}}, \bibinfo {author} {\bibfnamefont {M.}~\bibnamefont {Fujita}},
  \bibinfo {author} {\bibfnamefont {H.}~\bibnamefont {Ajiki}}, \ and\ \bibinfo
  {author} {\bibfnamefont {M.}~\bibnamefont {Sigrist}},\ }\href {\doibase
  10.1103/PhysRevB.59.8271} {\bibfield  {journal} {\bibinfo  {journal} {Phys.
  Rev. B}\ }\textbf {\bibinfo {volume} {59}},\ \bibinfo {pages} {8271}
  (\bibinfo {year} {1999})}\BibitemShut {NoStop}%
\bibitem [{\citenamefont {Fujimoto}\ \emph {et~al.}(2020)\citenamefont
  {Fujimoto}, \citenamefont {Koschke},\ and\ \citenamefont
  {Koshino}}]{PhysRevB.101.041112}%
  \BibitemOpen
  \bibfield  {author} {\bibinfo {author} {\bibfnamefont {M.}~\bibnamefont
  {Fujimoto}}, \bibinfo {author} {\bibfnamefont {H.}~\bibnamefont {Koschke}}, \
  and\ \bibinfo {author} {\bibfnamefont {M.}~\bibnamefont {Koshino}},\ }\href
  {\doibase 10.1103/PhysRevB.101.041112} {\bibfield  {journal} {\bibinfo
  {journal} {Phys. Rev. B}\ }\textbf {\bibinfo {volume} {101}},\ \bibinfo
  {pages} {041112} (\bibinfo {year} {2020})}\BibitemShut {NoStop}%
\bibitem [{\citenamefont {Zhang}\ \emph {et~al.}(2020)\citenamefont {Zhang},
  \citenamefont {Gao},\ and\ \citenamefont {Xiao}}]{zhang2020topological}%
  \BibitemOpen
  \bibfield  {author} {\bibinfo {author} {\bibfnamefont {Y.}~\bibnamefont
  {Zhang}}, \bibinfo {author} {\bibfnamefont {Y.}~\bibnamefont {Gao}}, \ and\
  \bibinfo {author} {\bibfnamefont {D.}~\bibnamefont {Xiao}},\ }\href@noop {}
  {\bibfield  {journal} {\bibinfo  {journal} {Physical Review B}\ }\textbf
  {\bibinfo {volume} {101}},\ \bibinfo {pages} {041410} (\bibinfo {year}
  {2020})}\BibitemShut {NoStop}%
\bibitem [{\citenamefont {Su}\ and\ \citenamefont
  {Lin}(2020)}]{su2020topological}%
  \BibitemOpen
  \bibfield  {author} {\bibinfo {author} {\bibfnamefont {Y.}~\bibnamefont
  {Su}}\ and\ \bibinfo {author} {\bibfnamefont {S.-Z.}\ \bibnamefont {Lin}},\
  }\href@noop {} {\bibfield  {journal} {\bibinfo  {journal} {Physical Review
  B}\ }\textbf {\bibinfo {volume} {101}},\ \bibinfo {pages} {041113} (\bibinfo
  {year} {2020})}\BibitemShut {NoStop}%
\bibitem [{\citenamefont {Thouless}(1983)}]{PhysRevB.27.6083}%
  \BibitemOpen
  \bibfield  {author} {\bibinfo {author} {\bibfnamefont {D.~J.}\ \bibnamefont
  {Thouless}},\ }\href {\doibase 10.1103/PhysRevB.27.6083} {\bibfield
  {journal} {\bibinfo  {journal} {Phys. Rev. B}\ }\textbf {\bibinfo {volume}
  {27}},\ \bibinfo {pages} {6083} (\bibinfo {year} {1983})}\BibitemShut
  {NoStop}%
\bibitem [{\citenamefont {Nakanishi}\ and\ \citenamefont
  {Ando}(2001)}]{nakanishi2001conductance}%
  \BibitemOpen
  \bibfield  {author} {\bibinfo {author} {\bibfnamefont {T.}~\bibnamefont
  {Nakanishi}}\ and\ \bibinfo {author} {\bibfnamefont {T.}~\bibnamefont
  {Ando}},\ }\href@noop {} {\bibfield  {journal} {\bibinfo  {journal} {Journal
  of the Physical Society of Japan}\ }\textbf {\bibinfo {volume} {70}},\
  \bibinfo {pages} {1647} (\bibinfo {year} {2001})}\BibitemShut {NoStop}%
\bibitem [{\citenamefont {Uryu}(2004)}]{PhysRevB.69.075402}%
  \BibitemOpen
  \bibfield  {author} {\bibinfo {author} {\bibfnamefont {S.}~\bibnamefont
  {Uryu}},\ }\href {\doibase 10.1103/PhysRevB.69.075402} {\bibfield  {journal}
  {\bibinfo  {journal} {Phys. Rev. B}\ }\textbf {\bibinfo {volume} {69}},\
  \bibinfo {pages} {075402} (\bibinfo {year} {2004})}\BibitemShut {NoStop}%
\bibitem [{\citenamefont {Slater}\ and\ \citenamefont
  {Koster}(1954)}]{PhysRev.94.1498}%
  \BibitemOpen
  \bibfield  {author} {\bibinfo {author} {\bibfnamefont {J.~C.}\ \bibnamefont
  {Slater}}\ and\ \bibinfo {author} {\bibfnamefont {G.~F.}\ \bibnamefont
  {Koster}},\ }\href {\doibase 10.1103/PhysRev.94.1498} {\bibfield  {journal}
  {\bibinfo  {journal} {Phys. Rev.}\ }\textbf {\bibinfo {volume} {94}},\
  \bibinfo {pages} {1498} (\bibinfo {year} {1954})}\BibitemShut {NoStop}%
\bibitem [{\citenamefont {Cao}\ \emph {et~al.}(2018{\natexlab{b}})\citenamefont
  {Cao}, \citenamefont {Fatemi}, \citenamefont {Demir}, \citenamefont {Fang},
  \citenamefont {Tomarken}, \citenamefont {Luo}, \citenamefont
  {Sanchez-Yamagishi}, \citenamefont {Watanabe}, \citenamefont {Taniguchi},
  \citenamefont {Kaxiras} \emph {et~al.}}]{cao2018correlated}%
  \BibitemOpen
  \bibfield  {author} {\bibinfo {author} {\bibfnamefont {Y.}~\bibnamefont
  {Cao}}, \bibinfo {author} {\bibfnamefont {V.}~\bibnamefont {Fatemi}},
  \bibinfo {author} {\bibfnamefont {A.}~\bibnamefont {Demir}}, \bibinfo
  {author} {\bibfnamefont {S.}~\bibnamefont {Fang}}, \bibinfo {author}
  {\bibfnamefont {S.~L.}\ \bibnamefont {Tomarken}}, \bibinfo {author}
  {\bibfnamefont {J.~Y.}\ \bibnamefont {Luo}}, \bibinfo {author} {\bibfnamefont
  {J.~D.}\ \bibnamefont {Sanchez-Yamagishi}}, \bibinfo {author} {\bibfnamefont
  {K.}~\bibnamefont {Watanabe}}, \bibinfo {author} {\bibfnamefont
  {T.}~\bibnamefont {Taniguchi}}, \bibinfo {author} {\bibfnamefont
  {E.}~\bibnamefont {Kaxiras}},  \emph {et~al.},\ }\href@noop {} {\bibfield
  {journal} {\bibinfo  {journal} {Nature}\ }\textbf {\bibinfo {volume} {556}},\
  \bibinfo {pages} {80} (\bibinfo {year} {2018}{\natexlab{b}})}\BibitemShut
  {NoStop}%
\bibitem [{\citenamefont {Nam}\ and\ \citenamefont
  {Koshino}(2017)}]{nam2017lattice}%
  \BibitemOpen
  \bibfield  {author} {\bibinfo {author} {\bibfnamefont {N.~N.}\ \bibnamefont
  {Nam}}\ and\ \bibinfo {author} {\bibfnamefont {M.}~\bibnamefont {Koshino}},\
  }\href@noop {} {\bibfield  {journal} {\bibinfo  {journal} {Physical Review
  B}\ }\textbf {\bibinfo {volume} {96}},\ \bibinfo {pages} {075311} (\bibinfo
  {year} {2017})}\BibitemShut {NoStop}%
\bibitem [{\citenamefont {Koshino}\ \emph {et~al.}(2018)\citenamefont
  {Koshino}, \citenamefont {Yuan}, \citenamefont {Koretsune}, \citenamefont
  {Ochi}, \citenamefont {Kuroki},\ and\ \citenamefont
  {Fu}}]{koshino2018maximally}%
  \BibitemOpen
  \bibfield  {author} {\bibinfo {author} {\bibfnamefont {M.}~\bibnamefont
  {Koshino}}, \bibinfo {author} {\bibfnamefont {N.~F.}\ \bibnamefont {Yuan}},
  \bibinfo {author} {\bibfnamefont {T.}~\bibnamefont {Koretsune}}, \bibinfo
  {author} {\bibfnamefont {M.}~\bibnamefont {Ochi}}, \bibinfo {author}
  {\bibfnamefont {K.}~\bibnamefont {Kuroki}}, \ and\ \bibinfo {author}
  {\bibfnamefont {L.}~\bibnamefont {Fu}},\ }\href@noop {} {\bibfield  {journal}
  {\bibinfo  {journal} {Physical Review X}\ }\textbf {\bibinfo {volume} {8}},\
  \bibinfo {pages} {031087} (\bibinfo {year} {2018})}\BibitemShut {NoStop}%
\bibitem [{\citenamefont {Koshino}(2015)}]{koshino2015interlayer}%
  \BibitemOpen
  \bibfield  {author} {\bibinfo {author} {\bibfnamefont {M.}~\bibnamefont
  {Koshino}},\ }\href@noop {} {\bibfield  {journal} {\bibinfo  {journal} {New
  Journal of Physics}\ }\textbf {\bibinfo {volume} {17}},\ \bibinfo {pages}
  {015014} (\bibinfo {year} {2015})}\BibitemShut {NoStop}%
\bibitem [{\citenamefont {Koshino}\ and\ \citenamefont
  {Moon}(2015)}]{koshino2015electronic}%
  \BibitemOpen
  \bibfield  {author} {\bibinfo {author} {\bibfnamefont {M.}~\bibnamefont
  {Koshino}}\ and\ \bibinfo {author} {\bibfnamefont {P.}~\bibnamefont {Moon}},\
  }\href@noop {} {\bibfield  {journal} {\bibinfo  {journal} {Journal of the
  Physical Society of Japan}\ }\textbf {\bibinfo {volume} {84}},\ \bibinfo
  {pages} {121001} (\bibinfo {year} {2015})}\BibitemShut {NoStop}%
\bibitem [{\citenamefont {Weckbecker}\ \emph {et~al.}(2016)\citenamefont
  {Weckbecker}, \citenamefont {Shallcross}, \citenamefont {Fleischmann},
  \citenamefont {Ray}, \citenamefont {Sharma},\ and\ \citenamefont
  {Pankratov}}]{weckbecker2016low}%
  \BibitemOpen
  \bibfield  {author} {\bibinfo {author} {\bibfnamefont {D.}~\bibnamefont
  {Weckbecker}}, \bibinfo {author} {\bibfnamefont {S.}~\bibnamefont
  {Shallcross}}, \bibinfo {author} {\bibfnamefont {M.}~\bibnamefont
  {Fleischmann}}, \bibinfo {author} {\bibfnamefont {N.}~\bibnamefont {Ray}},
  \bibinfo {author} {\bibfnamefont {S.}~\bibnamefont {Sharma}}, \ and\ \bibinfo
  {author} {\bibfnamefont {O.}~\bibnamefont {Pankratov}},\ }\href@noop {}
  {\bibfield  {journal} {\bibinfo  {journal} {Physical Review B}\ }\textbf
  {\bibinfo {volume} {93}},\ \bibinfo {pages} {035452} (\bibinfo {year}
  {2016})}\BibitemShut {NoStop}%
\bibitem [{\citenamefont {Hatsugai}\ and\ \citenamefont
  {Fukui}(2016)}]{hatsugai2016bulk}%
  \BibitemOpen
  \bibfield  {author} {\bibinfo {author} {\bibfnamefont {Y.}~\bibnamefont
  {Hatsugai}}\ and\ \bibinfo {author} {\bibfnamefont {T.}~\bibnamefont
  {Fukui}},\ }\href@noop {} {\bibfield  {journal} {\bibinfo  {journal}
  {Physical Review B}\ }\textbf {\bibinfo {volume} {94}},\ \bibinfo {pages}
  {041102} (\bibinfo {year} {2016})}\BibitemShut {NoStop}%
\end{thebibliography}%

\end{document}